\documentclass [11pt]{revtex4}    
\usepackage{amssymb,epsf}
\usepackage{latexsym}
\begin{document}
\title{Boundary Conditions in Stepwise Sine-Gordon Equation and Multi-Soliton Solutions}
\author{N. Riazi
\footnote{email: riazi@physics.susc.ac.ir} and A. Sheykhi
\footnote{email: sheykhi@gmail.com}}
\address{Physics Department and Biruni Observatory,
Shiraz University, Shiraz 71454, Iran}

\begin{abstract}
We study the stepwise sine-Gordon equation, in which the system
parameter is different for positive and negative values of the
scalar field. By applying appropriate boundary conditions, we
derive relations between the soliton velocities before and after
collisions. We investigate the possibility of formation of heavy
soliton pairs from light ones and vise versa. The concept of
soliton gun is introduced for the first time; a light pair is
produced moving with high velocity, after the annihilation of a
bound, heavy pair. We also apply  boundary conditions to static,
periodic and quasi-periodic solutions.
\end{abstract} \maketitle
\section{Introduction} \label{sec1}
  The Sine-Gordon equation (SGE) is a nonlinear equation which is encountered in various
  contexts \cite{RGH}. As an example, is can be shown that the classical string on a two-sphere is more or
less equivalent to the sine-Gordon model \cite{poh}. Mikhailov
\cite{mik} has  considered a system which can be understood as a
T-dual to the classical string on a two-sphere. He has shown that
there is a projection map from the phase space of this model to
the phase space of the sine-Gordon model.

One traditional way to obtain SGE in 1+1 dimensions is in the
context of differential  geometry  and the theory of  surfaces.
The fundamental forms  of a surface are not independent of each
other and must satisfy  the Gauss-Codazzi  equations. It can be
easily shown that the Gauss-Codazzi equations lead to the
Sine-Gordon equation for a surface of constant negative curvature.
The spatial shapes of the surfaces corresponding to the one and
two soliton solutions have been shown in \cite{TU}.

The SGE  can be written in the following form, using an
appropriate choice of coordinates:
\begin{equation}\label{SGE2}
\frac{\partial^2 \varphi}{\partial x^2}-\frac{\partial^2
\varphi}{c^2\partial t^2}=a\sin \varphi,
\end{equation}
where $a$ is a constant parameter. In this form, SGE is a
relativistic equation, which can be obtained from the following
Lagrangian density:
\begin{equation}\label{lag}
{\cal L}=\frac{1}{2}\partial^ \mu \varphi \partial_{ \mu}
\varphi-a( 1-\cos\varphi).
\end{equation}
The energy momentum tensor can be obtained from the Noether's
theorem \cite{gud}:
\begin{equation}\label{TM}
 T^ {\mu\nu}= \partial^\mu
\varphi\partial^\nu \varphi-\eta^{\mu\nu}{\cal L},
\end{equation}
where $\eta^{\mu\nu}$ is  the Minkowski metric for the 1+1
dimensional spacetime. The SGE  is an integrable equation. An
integrable system has as many constants of motion as the number of
the system's degrees of freedom \cite{Das}. Since the SGE  have
infinite degrees of freedom (as in any other continuous medium),
its integrability means that it has infinitely many constants of
motion. Integrable equations have peculiar solutions  named
solitons (or kinks in the case of SGE). Solitons are localized
waves  which do not disperse as they move in the nonlinear medium.
They also preserve their shape after the collision with other
solitons or localized inhomogeneities in the medium. It is usually
claimed that a soliton reobtains its initial velocity after
collision with  another soliton and  only a phase shift results
form the collision \cite{Raj}. For the system we will investigate
in this paper,  we will show  that this claim is indeed not
correct.

The one soliton solution (kink) of (\ref{SGE2}) is
\begin{equation}\label{kink}
\varphi(x,t)=4\arctan {e^{\gamma a(x-vt)}}.
\end{equation}
 This solution can be obtained by different means (e.g via Backlund
 transformations or by direct integration and applying a Lorentz boost). The effective width
 of a kink at rest is about $a^{-1}$.

The total energy and linear momentum of a single soliton (kink) of
the SGE can be obtained from the following relations:
\begin{equation}
E= \int{T^0_{0}}dx=\gamma Mc^2 ~~~~ P=\int{T^1_{0}}dx= \gamma Mv.
\end{equation}
Here, $\gamma$ is related to the kink velocity via the $\gamma =
\frac{1}{\sqrt{1-\frac{v^2}{c^2}}}$, and the kink's rest energy
(mass) is given by $M = 8\frac{\sqrt{a}}{c^2}$.

For one soliton solutions, it can be shown that:
\begin{equation}\label{pen}
E^2=P^2c^2+ M^2c^4.
\end{equation}
In this respect and other respects to be discussed, the kink
behaves like a classical relativistic  particle, although it is an
extended object and has wave nature inherent in it. Many
applications have been found for 1+1 dimensional SGE (see
\cite{RGH} for a review). Some generalizations of this equation
can be found in \cite{RAZ} and \cite{RM}.

 The organization of this paper is as follows. In section \ref{sec2}
we will introduce the stepwise sine-Gordon equation in which the
parameter $a$ is different for negative and positive values of the
scalar field $\varphi$. By applying appropriate boundary
conditions, we derive relations between the kink velocities before
and after interactions. We investigate the possibility of
formation of heavy soliton pairs from light ones and vise versa.
In section \ref{sec4}, we briefly study the relation between the
simplex structure and topological charges of different solutions.
In section \ref{static}, we apply  boundary conditions to static,
periodic and quasi-periodic solutions. Section \ref{sec5} is
devoted to summary and conclusions.

\section{Stepwise SGE in 1+1 dimensions and Boundary Conditions}\label{sec2}
By the stepwise SGE, we mean imposing the following condition on
equation (\ref{SGE2}):
\begin{eqnarray}\label{cond}
a=a_{1},~~ {\rm for}~~ \varphi < 0 \\
a=a_{2},~~ {\rm for}~~ \varphi > 0.
\end{eqnarray}
For each of the regions $\varphi >0 $ and  $\varphi <0 $, the
scalar field $\varphi$ satisfies (\ref{SGE2}) with the
corresponding value of the parameter $a$. The region needing
particular attention is the boundary region separating the
$\varphi <0$ and $\varphi >0$ regions. If we have $\varphi<0 $ for
$t<0$ and $\varphi>0$ for $t>0$ (or vice versa),  we can show that
the following boundary condition holds:
\begin{equation}\label{jun1}
\forall x ;~~~ \frac{ \partial \varphi}{\partial t}\mid_{0^+}=
\frac{\partial \varphi}{\partial t}\mid_{0^-}.
\end{equation}
In order to arrive at this boundary condition, we can integrate
the dynamical equation from $t=-\epsilon$ to $t=+\epsilon$ and let
$\epsilon \rightarrow 0$. In a similar way, if the two regions
$\varphi>0 $ and $\varphi<0$ connect to each other at $x=0$, we
obtain the following junction condition:
\begin{equation}\label{jun2} \forall t;~~~ \frac{
\partial \varphi}{\partial x}\mid_{0^+}= \frac{\partial
\varphi}{\partial x}\mid_{0^-}.
\end{equation}
This boundary condition is obtained by integrating the dynamical
equation over $x$ from $x=-\epsilon$ to $+\epsilon$ and letting
$\epsilon \rightarrow 0$.  These conditions provide the key
relations for studying the dynamics of two soliton interactions in
the stepwise SGE.

Like the conventional SG equation, the stepwise SGE is a
relativistically covariant equation. One expects the boundary
conditions mentioned above to be expressible in a manifestly
covariant form. This is indeed the case. In deriving equations
(\ref{jun1}) and (\ref{jun2}), we have assumed that the boundary
between $\varphi <0$ and $\varphi >0$ regions occur at $t=0$ or
$x=0$ hyper-surfaces. In the  examples which will follow, this
assumption is actually fulfilled. However, in an arbitrary frame
of reference, where the boundary between these two regions occurs
at $x_b(t)$, the junction conditions should be written in the
covariant form
\begin{equation}
\left( \frac{\partial \varphi}{\partial x^\mu}\right)_-=\left(
\frac{\partial \varphi}{\partial x^\mu}\right)_+,
\end{equation}
applied at $x=x_b(t)$ (the - and + signs correspond to either
sides of the boundary).

\subsection{Free Kink-Antikink Collision}\label{subs1}
\begin{figure}[t]
 \epsfxsize=10cm
 \centerline{\epsffile{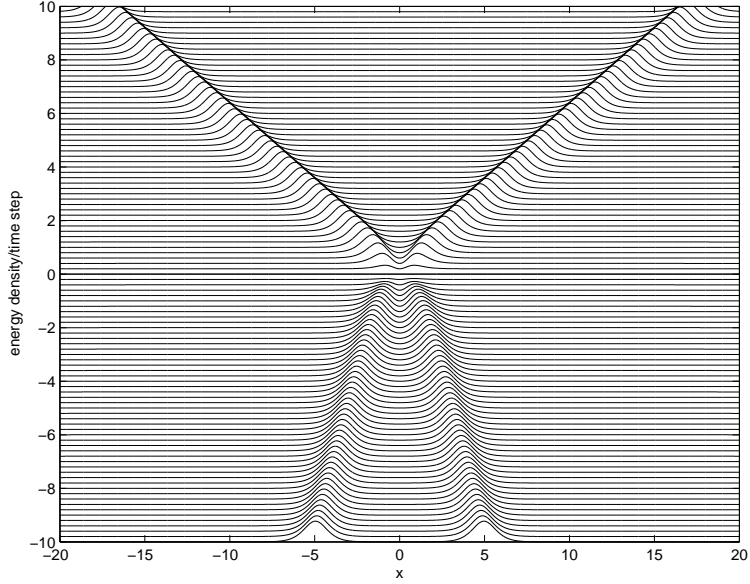}}
           \caption{Annihilation of a free, heavy kink-anti-kink pair
           and the
           production of a light pair in the stepwise SGE.}
           \label{fig1}
           \end{figure}

 The free kink-anti-kink solution of the SGE
for $m^2_{i}>1$ is \cite{lamb}:
\begin{equation}\label{kak1}
\varphi(x,t)=-4\arctan\left(\frac{m_{i}}{\sqrt{m^2_{i}-1}}
\frac{\sinh(\sqrt{(m^2_{i}-1)a_{i}}ct)}{\cosh(m_{i}\sqrt{a_{i}}x)}\right)
\end{equation}
This solution holds for $t>0$ (with \textit{i}=1) and $t<0$ (with
\textit{i}=2). If we write the suitable parameter for each
interval, namely
\begin{eqnarray}\label{kak2}
 a=a_{1}~~~ {\rm and}~~m=m_{1}~~~ {\rm for}~~~ t>0\\
 a=a_{2}~~~ {\rm and}~~m=m_{2}~~~ {\rm for}~~~ t<0
 \end{eqnarray}
in the SGE in each region, then by use the first boundary
condition (\ref{jun1}) we easily obtain
\begin{equation}\label{kak2}
m_{1}\sqrt{a_{1}}=m_{2}\sqrt{a_{2}}
\end{equation}
It can be shown that $m_{i}$ is related to the soliton's velocity
far from the interaction region via
\begin{equation}\label{kak3}
m_{i}=\gamma_{i}=\frac{1}{\sqrt{1-\frac{v^2_{i}}{c^2}}}
\end{equation}
And therefore using the relation $ M_{i}c^2=8\sqrt{a_{i}}$, we can
rewrite equation (\ref{kak2}) as
\begin{equation}\label{kak4}
\gamma_{1}M_{1}c^2 = \gamma_{2}M_{2}c^2
\end{equation}
This equation shows the energy conservation in  annihilation  of a
free kink-anti-kink  pair and  the creation of another one with
different masses. This process has been shown in Figure
\ref{fig1}.
\subsection{Bound State Kink-Anti-Kink Pair (Breather Solution)}\label{subs2}
\begin{figure}[t]
 \epsfxsize=10cm
 \centerline{\epsffile{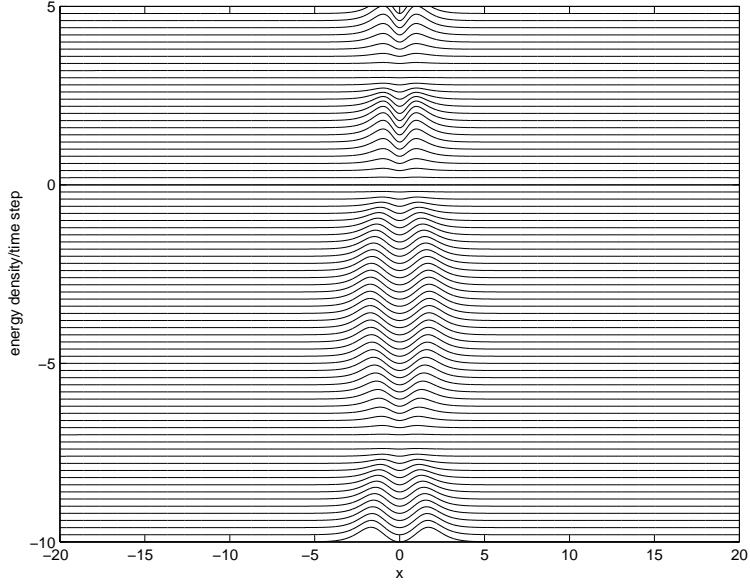}}
           \caption{The breather solution in the stepwise SGE.}
           \label{fig2}
           \end{figure}

\begin{figure}[t]
 \epsfxsize=10cm
 \centerline{\epsffile{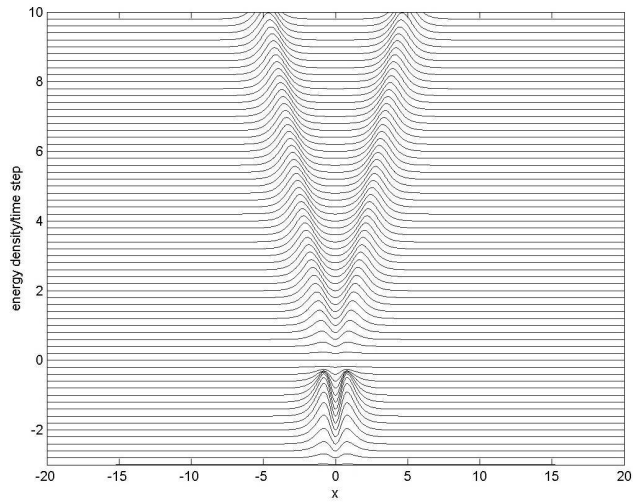}}
           \caption{Annihilation of a heavy breather and creation of a light
           free pair.}
           \label{fig3}
           \end{figure}

The kink-anti-kink solution for $m^2_{i}<1$ is
\begin{equation}\label{bkak}
\varphi(x,t)=\pm4\arctan\left(\frac{m_{i}}{\sqrt{1-m^2_{i}}}
\frac{\sinh(\sqrt{(1-m^2_{i})a_{i}}ct)}{\cosh(m_{i}\sqrt{a_{i}}x)}\right).
\end{equation}
From the sign of $ \varphi$, we can distinguish two regions:
\begin{eqnarray}\label{bkak2}
 ~~ a=a_{1}~~~ m=m_{1}~~~ {\rm for}~~~-\tau_{1}<t<0\\
  a=a_{2}~~~ m=m_{2}~~~    {\rm for}~~~     0<t<\tau_{2}
\end{eqnarray}
where
\begin{equation}\label{bkak3}
\tau_{1}=\frac{\pi}{\sqrt{(1-m^2_{1})a_{1}}c}~~~~
\tau_{2}=\frac{\pi}{\sqrt{(1-m^2_{2})a_{1}}c}.
\end{equation}
Imposing the boundary condition (\ref{jun1}) we get:
\begin{equation}\label{bkak4}
m_{1}\sqrt{a_{1}}=m_{2}\sqrt{a_{2}}
\end{equation}
Which leads to equation (\ref{kak4}). From  equation
(\ref{bkak3}), we can write
\begin{equation}\label{bkak5}
\frac{\tau_{2}}{\tau_{1}}=\frac{a_{1}}{a_{2}}
\sqrt{\frac{1-m^2_{1}}{1-\left(\frac{a_{1}}{a_{2}}\right)^2m^2_{1}}}.
\end{equation}
It is clear that we have no breather in the case $m_{2}>1$ and $
t>0$. In this case, the breather will decay to free light pairs.
This case is illustrated in Figures \ref{fig2} and \ref{fig3}. So,
we distinguish two
cases for $a_{2}<a_{1}$: \\


\begin{itemize}
\item{ $m_{1}<\frac{a_{2}}{a_{1}}$. In this case one breather
turns into another breather with a different oscillation period.}
\item{ $m_{1}>\frac{a_{2}}{a_{1}}$. In this case one breather
decays into a free one  with lighter  rest mass. Let us coin this
phenomenon, which is the most important result of this work, as
the soliton gun (See Fig. 3). This phenomenon is apparently
similar, to the annihilation of $\mu^+\mu^-$ bound pair and the
creation of a free $e^+ e^-$ pair.} \end{itemize}

 By suitable choice of initial
conditions, it is possible to create a heavy breather via light
kink-anti-kink collision. This breather then decays back to the
initial free pair. This is similar to the \textit{resonance}
phenomenon in elementary particle physics.
\subsection{Kink-Kink Collision}\label{subs3}
\begin{figure}[t]
 \epsfxsize=10cm
 \centerline{\epsffile{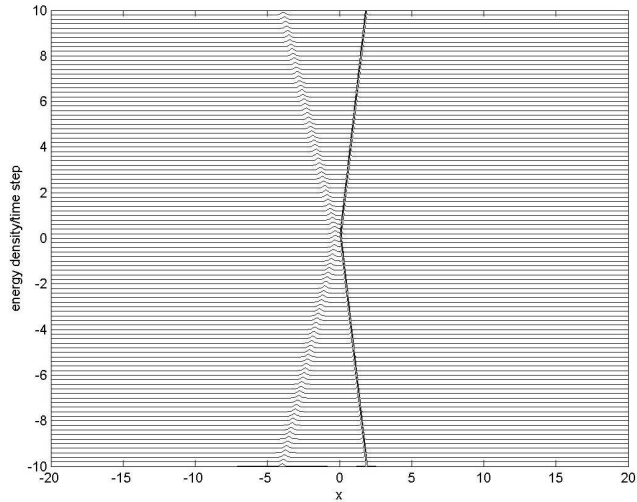}}
           \caption{Collision between two kinks with different rest masses in the stepwise SGE. The
           RHS kink is 5 times as massive as the LHS one.}
           \label{fig4}
           \end{figure}

This case corresponds to $m^2_{i}>1$. Using the kink-kink solution
\begin{equation}\label{kk}
\varphi(x,t)=4\arctan\left(\frac{\sqrt{m^2_{i}-1}}{m_{i}}
\frac{\sinh(m_{i}\sqrt{a_{i}}x)}{\cosh(\sqrt{(m^2_{i}-1)a_{i}}ct)}\right),
\end{equation}
and applying the boundary condition (\ref{jun2}), after a little
calculation we obtain
\begin{equation}\label{kkj}
\sqrt{(m^2_{1}-1)a_{1}}= \sqrt{(m^2_{2}-1)a_{2}}
\end{equation}
or
\begin{equation}\label{kkj2}
\gamma_{1}M_{1}v_{1}=\pm \gamma_{2}M_{2}v_{2}
\end{equation}
After the collision, the velocity  of each kink will change (see
Figure \ref{fig4}). This is in contradiction with the old idea
that solitons will  reobtain their initial velocities after
collision and only a time delay results. Since our system is
relativistic, we can consider a reference frame observer in which
one of the solitons is at rest before  the collision and acquires
a velocity after the collision. This is similar to the collision
of a
billiard ball to another ball initially at rest. \\

\begin{figure}[t]
 \epsfxsize=10cm
 \centerline{\epsffile{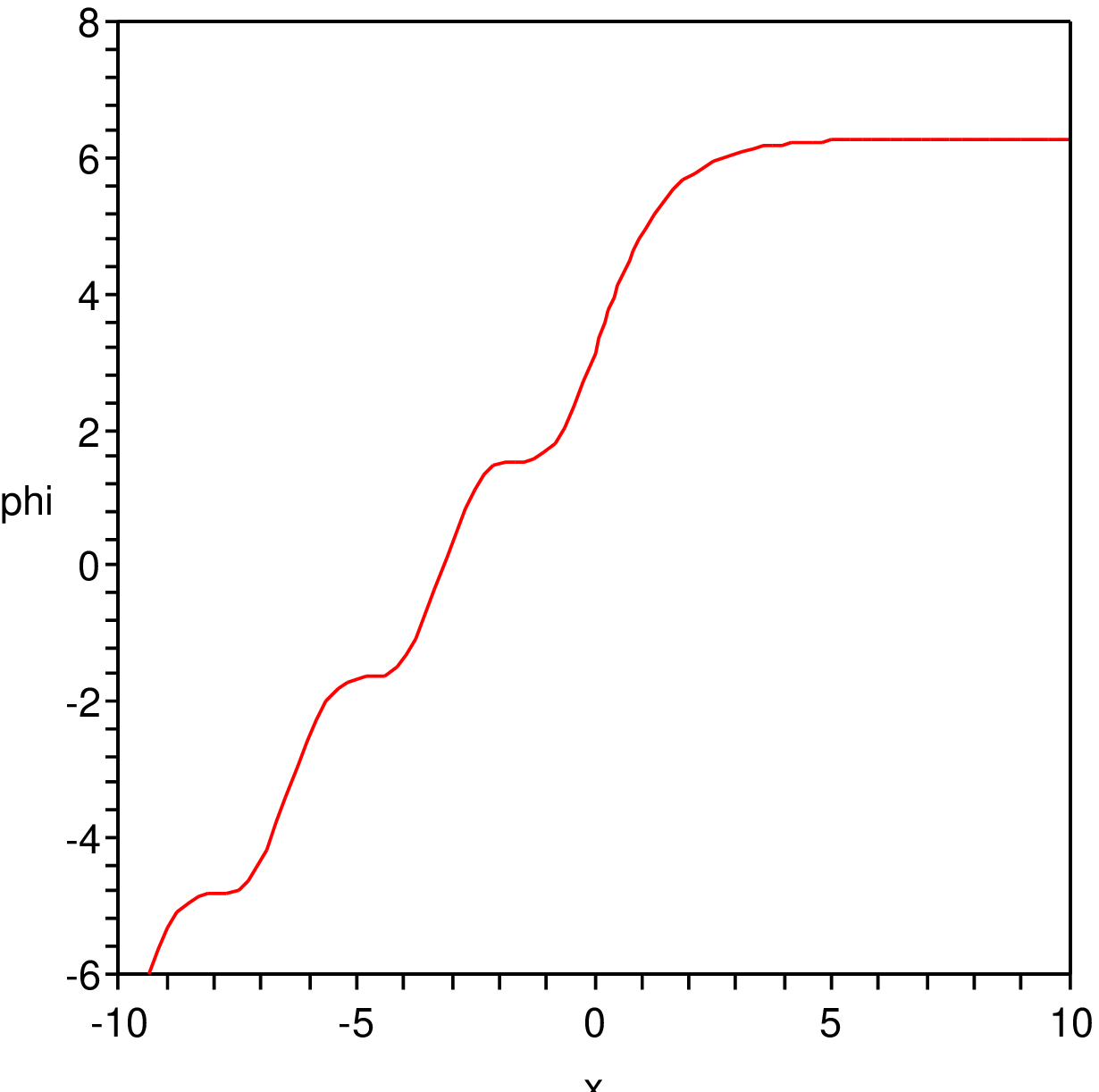}}
           \caption{Static solution corresponding to a kink chain, smoothly joining to a kink in the stepwise SGE.}
           \label{chainkink}
           \end{figure}

\section{Simplex structure and topological charge for the stepwise SGE}\label{sec4}
Simplexes are r-dimensional geometrical structures, which are
shown  by the symbol $\sigma_{r}= <p_{0}p_{1}...p_{r}>$. For
example, a 0-simplex $<p_{0}>$ is  a point or a vertex, and a
1-simplex $<p_{0}p_{1}>$ is a line or an edge. A 2-simplex
$<p_{0}p_{1}p_{2}>$ is defined to be a triangle with its interior
included and etc. The set of finite number of simplexes in $R^m$
which are properly fitted together is called simplical complex and
denoted by $K$.
 By 'properly', we mean that:\\
i) An arbitrary face of a simplex of $K$ belongs to $K$, that is,
if $\sigma\in K$ and $\sigma'\leq \sigma$ then $\sigma'\in K$.\\
ii) If $\sigma$ and $\sigma'$ are two simplexes of $K$, the
intersection $\sigma\cap\sigma'$ is either \textit{empty} or a
\textit{face} of $\sigma$ and $\sigma'$; that is, if $\sigma,
\sigma'\in K$ then either $\sigma'\cap \sigma ={\O}$ or
$\sigma\cap\sigma'\leq\sigma$ and $\sigma\cap
\sigma'\leq\sigma'$.\\
 If we assign \textit{orientation} to an
r-simplex for $r\geq1$, it is called \textit{r-chain} and is shown
with $\sigma_{r}=(p_{0}p_{1}...p_{r})(r>0)$.\\ A \textit{boundary}
$\partial_{r}\sigma_{r}$  of $\sigma_{r}$ is an $(r-1)$ -chain
defined by:
\begin{equation}\label{bop}
\partial_{r}(p_{0}p_{1}...p_{r})\equiv \Sigma(-1)^i(p_{0}...\hat{p_{i}}...p_{r})
\end{equation}
where the point $p_{i}$ under 'hat'  is omitted. The
$\partial_{r}$ is called the  \textit{boundary operator}. The
boundary of an r-chain is a set of $(r-1)$-chains.  We can also
define \textit{r-cycle}, which is an r-dimensional orientated
complexes which has no boundary namely if
 $c$ is a r-cycle, then $\partial_{r}c=0$ \cite{naka}. \\
We can consider each soliton of the stepwise sine-Gordon equation
as a 1-chain in $\varphi$ space. The classical vacuum of the
potential of the system is $\varphi_{n}=2n\pi,~~n\in Z$, which are
discrete  points. The kink solution is an oriented simplex
 or chain,$(\varphi_{n}\varphi_{n+1})$ and the anti-kink can be shown
as a chain in the form
$(\varphi_{n+1}\varphi_{n})=-(\varphi_{n}\varphi_{n+1})$. If we
act the boundary operator on the one soliton solution (kink), we
obtain
\begin{equation}\label{bop2}
\partial
(\varphi_{n}\varphi_{n+1})=\varphi_{n+1}-\varphi_{n}=2\pi.
\end{equation}
$J^\mu=\frac{1}{2\pi}\epsilon^{\mu\nu}\partial_\nu \varphi$ is the
topological current which is identically conserved ($\partial^\mu
J_\mu =0$). The topological charge in the sine-Gordon system can
be written as:
\begin{equation}\label{chrg}
Q=\int J^o dx=
\frac{1}{2\pi}\left[\varphi(+\infty)-\varphi(-\infty)\right]
=~\frac{1}{2\pi}\partial_{1}~\left(\varphi(+\infty),\varphi(-\infty)\right).
\end{equation}
It is clear that topological charge can be obtained directly by
applying  the boundary operator on the soliton-chain. The
topological charge for the different solutions discussed in
pervious sections are as follows. For kink $(Q=+1)$, anti-kink
$(Q=-1)$, kink-kink $(Q=+2)$, breather and free kink-anti-kink
$(Q=0)$. The topological charge is
quantized and is independent of the $a_{i}$ parameters in the stepwise SGE.\\
In a similar way, we can consider multi-soliton solutions as
complexes in the scalar field $\varphi$. If we define the
sine-Gordon equation on  $S^1$ instead of $R$,  the single
valuedness condition of $\varphi$ implies that there could be no
single soliton solutions, and the multi soliton solutions are as
\textit{1-cycle}s. It is natural that the total charge of the
multi soliton solutions on a compact manifold is zero
$(Q=\frac{1}{2\pi}
\partial_{1}c=0)$. In other words, the total topological charge of
the sine-Gordon equation (ordinary or stepwise), on the compact
space $S^1$ is always zero.

\begin{figure}[t]
 \epsfxsize=10cm
 \centerline{\epsffile{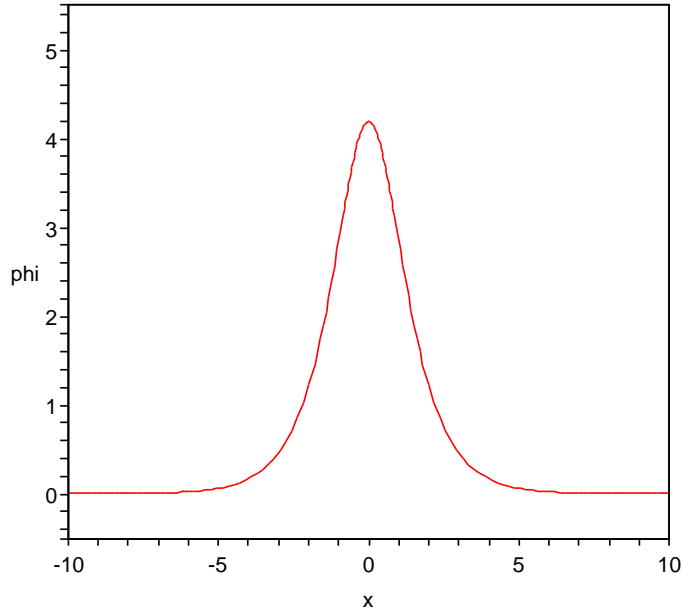}}
           \caption{Static solution corresponding to a kink, smoothly joining to a part of the kink chain and then
           to an anti-kink in the stepwise SGE.}
           \label{kinkperant}
           \end{figure}

\section{Static solutions}\label{static}
It is well-known that the SG system has also static, periodic and
quasi-periodic solutions which can be interpreted as a chain of
kinks and anti-kinks or a chain of kinks (anti-kinks). These are
given by Jacobi elliptic functions. Consider the first integral of
the static SG equation:
\begin{equation}
\frac{1}{2}(\frac{d\varphi}{dx})^2+a\cos(\varphi)=C,
\end{equation}
where $C$ is an integration constant. For single soliton
solutions, it is easily seen that $C=a$ due to boundary conditions
at $x\rightarrow \pm \infty$. For $C<a$, the static solutions can
be expressed in terms of  the Jacobi SN functions \cite{ba}:
\begin{equation}
\varphi(x)=(2n+1)\pi\pm 2\arcsin(kSN(\sqrt{a}(x-x_1),k)),
\end{equation}
where $k$ and $x_1$ are integration constants, and $n$ is an
integer. This solution oscillates around $\varphi=\pi$, which
corresponds to a lattice of alternating kinks and anti-kinks. As
$k\rightarrow 1$, the distance between neighboring kinks and
anti-kinks increases, and in the limit $k=1$, we will have an
isolated kink (or anti-kink) solution.

\begin{figure}[t]
 \epsfxsize=10cm
 \centerline{\epsffile{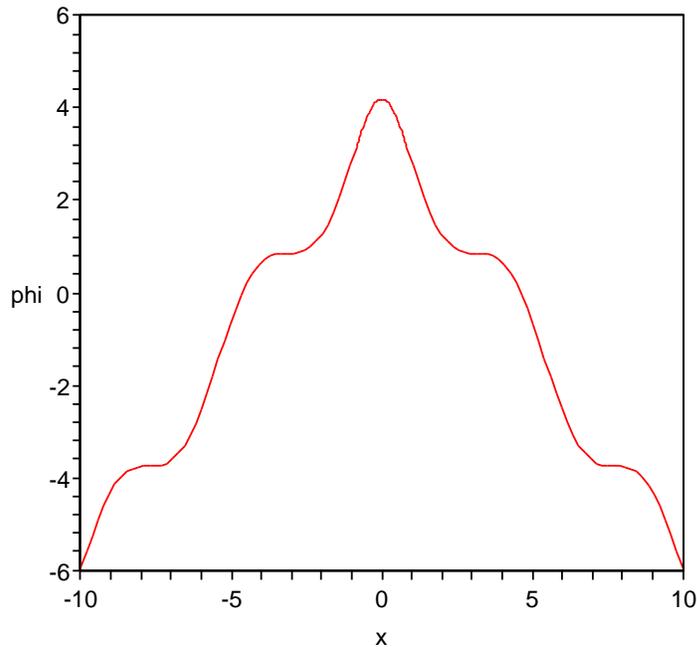}}
           \caption{Static solution corresponding to a kink chain, smoothly joining to an anti-kink chain
            in the stepwise SGE.}
           \label{chainperchain}
           \end{figure}

\begin{figure}[t]
 \epsfxsize=10cm
 \centerline{\epsffile{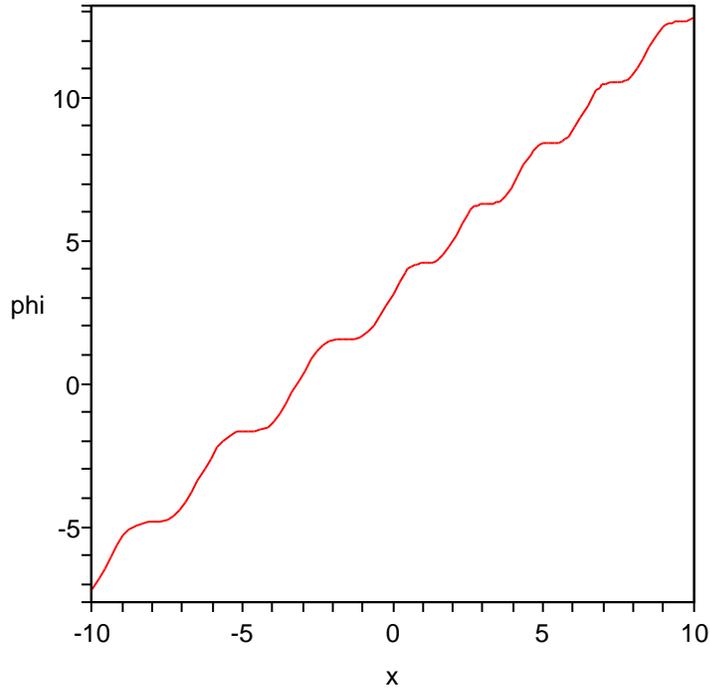}}
           \caption{Static solution corresponding to a kink chain, smoothly joining to another kink chain
            in the stepwise SGE.}
           \label{chainchain}
           \end{figure}

For $C>a$, we will have the following static solution:
\begin{equation}
\varphi(x)=(2n+1)\pi\pm x\pm 2\arcsin(kSN(\sqrt{a}(x-x_0),k)).
\end{equation}
This solution corresponds to a lattice of kinks, with the distance
between the kinks  increasing as $k\rightarrow 1$.

Based on the fore-mentioned solutions, we now turn to the stepwise
SG equation with \begin{equation} a=a_1 \ \ \ {\rm if} \ \ \
\varphi<\pi,
\end{equation}
and
\begin{equation}a=a_2 \ \ \ {\rm if} \ \ \
\varphi>\pi. \end{equation}

We must now apply the boundary conditions at $x_j$, where $x_j$'s
are the solutions of $\varphi(x_j)=\pi$. We can distinguish
several categories of static solutions for the stepwise SG
equation. Among different possibilities, let us introduce the
following four cases:
\begin{itemize} \item Kink lattice, attached to a single
kink. The boundary condition leads to:
\begin{equation}
1+2k_1\sqrt{a_1}=2\sqrt{a_2}, \ \ (k_2=1).
\end{equation}

A sample solution is shown in Figure \ref{chainkink}.

\item Kink, smoothly attached to a part of the kink-antikink chain
and then to an anti-kink (Figure \ref{kinkperant}). The boundary
condition leads to
\begin{equation}
\sqrt{a_1}=k_2\sqrt{a_2}, \ \ (k_1=1).
\end{equation}
This is an interesting solution, since it is a non-topological,
static  soliton with zero topological charge. It is well-known
that the conventional SG system does NOT have a static, zero
charge (non-topological) solution.
 \item Kink chain, smoothly attached to a part of the kink-antikink chain and then to an anti-kink chain
 (Figure \ref{chainperchain}). The boundary condition leads to
 \begin{equation}
 1+2k_1\sqrt{a_1}=2k_2\sqrt{a_2}.
 \end{equation}

\item Kink chain attached smoothly to another kink chain (see
Figure \ref{chainchain}):
\begin{equation}
k_1\sqrt{a_1}=k_2\sqrt{a_2}.
\end{equation}

\end{itemize}
\section{ summary and conclusion}\label{sec5}
In this paper, we studied the stepwise sine-Gordon system in which
the mass parameter '$a$' is different for positive and negative
values of the scalar field.  By applying appropriate boundary
conditions at the boundary between the two regions, which are
derived from the field equation, we extracted the relation between
the soliton velocities before and after the collision.

Similar to the boundary conditions in electrodynamics or in the
Schr\"odinger equation, the boundary conditions of the stepwise
SGE are obtained by integrating the field equation across the
boundary between the $\varphi <0$ and $\varphi >0$ regions. These
relations are consistent with energy and momentum conservation
relations for the system under consideration. We considered
different cases and showed that in such a system, it is possible
to transform a heavy pair of solitons to a light pair and vise
versa. The concept of soliton gun was introduced for the first
time. In this process, a heavy bound pair annihilates into a free
low mass pair, moving at high velocities. Since the kinks
belonging to different ($\varphi <0$ and $\varphi>0$) sectors have
different rest energies (masses), they cannot retrieve their
initial velocities after the collisions. The stepwise SGE is
therefore an interesting system for  studying the collisions
between solitons of different rest energies.

The similarity of the kinks and classical particles is further
demonstrated in the framework of the stepwise SGE. We emphasize
that in this system, the initial soliton velocities are not
necessarily retrieved after the collisions.

The application of boundary conditions to static, periodic and
quasi-periodic solutions was also worked out, leading to new
results, not present in the conventional SG system.
 \acknowledgements
 N. Riazi acknowledges the support of Shiraz University. Authors
 thank the referees for their very helpful comments and
 suggestions.


\end{document}